\documentclass[10pt,conference]{IEEEtran}
\usepackage{verbatim,amssymb,amsmath,graphicx}

\usepackage[T1]{fontenc}
\usepackage[latin9]{inputenc}
\usepackage{flushend}

\newcommand{\RR}{\mathbb{R}}
\newcommand{\ZZ}{\mathbb{Z}}

\newcommand{\Sc}{\mathcal{S}}
\newcommand{\Vc}{\mathcal{V}}

\newcommand{\cv}{\mathbf{c}}
\newcommand{\rv}{\mathbf{r}}
\newcommand{\sv}{\mathbf{s}}
\newcommand{\uv}{\mathbf{u}}
\newcommand{\vv}{\mathbf{v}}
\newcommand{\wv}{\mathbf{w}}
\newcommand{\xv}{\mathbf{x}}
\newcommand{\yv}{\mathbf{y}}
\newcommand{\zv}{\mathbf{z}}

\newcommand{\vol}{{\rm Vol}}

\newtheorem{example}{Example}
\newtheorem{defn}{Definition}

\begin{document}
\title{Secrecy Gain: a Wiretap Lattice Code Design}
\author{
\authorblockN{Jean-Claude Belfiore}
\authorblockA{Department of Communications and Electronics \\
TELECOM ParisTech\\
Paris\\
France\\
Email: belfiore@telecom-paristech.fr}
\and
\authorblockN{Fr\'ed\'erique Oggier}
\authorblockA{Division of Mathematical Sciences \\
School of Physical and Mathematical Sciences\\
Nanyang Technological University \\
Singapore\\
Email: frederique@ntu.edu.sg}
}
\maketitle

%
%

\begin{abstract}
We propose the notion of secrecy gain as a code design criterion for wiretap 
lattice codes to be used over an additive white Gaussian noise channel. Our 
analysis relies on the error probabilites of both the legitimate user and the 
eavesdropper. We focus on geometrical properties of lattices, described  
by their theta series, to characterize good wiretap codes.
\end{abstract}

%
%
\section{Introduction}

The wiretap channel was introduced by Wyner \cite{W-75} as a discrete
memoryless broadcast channel where the sender, Alice, transmits confidential
messages to a legal receiver Bob, in the presence of an eavesdropper Eve.
Wyner defined the perfect secrecy capacity as the maximum amount of 
information that Alice can send to Bob while insuring that Eve gets a 
negligeable amount of information. He also described a generic coding strategy
known as coset coding. While coset coding has been used in many coding 
scenarios (for ex. \cite{ZSE-02,PR-05}), Wyner used it to encode both data 
and random bits to confuse the eavesdropper. 
A more precise coset coding technique, called wiretap
II codes, was presented in \cite{OW-84}, where Alice enjoys a noiseless 
channel while Eve has to deal with a channel with erasures.
The question of determining the secrecy capacity of many classes of channels
has been addressed extensively recently, yielding a plethora of information
theoretical results on secrecy capacity.

There is a sharp contrast with the situation of wiretap code designs, where
very little is known. The most exploited approach to get practical codes so
far has been to use LDPC codes (for example \cite{TDCMM-07} for binary erasure 
and symmetric channels, \cite{KHMBK-98} for Gaussian channels).
We also note that wiretap II codes have been extended to more general settings
such as network coding in \cite{ES-07}. Finally, lattice codes for Gaussian
channels have been considered from an information theoretical point of view
in \cite{HY-09}.

The problem that we address in this paper is to propose a design 
criterion for constructing explicit lattice codes (of possibly small length)
to be used over additive white Gaussian noise channels. Assuming that Eve's 
channel is worse than the one of Alice, we analyse the probability of both
users to make a correct decision, and exhibit geometrical lattice properties
that maximize Alice's probability of making the right decision, while
minimizing Eve's probability of decoding successfully. These properties are 
captured by the theta series of the lattice used for encoding, which in turn 
is used to define the notion of {\em secrecy gain} as a measure of secrecy 
brough by the lattice wiretap codes. 
Note that we do not consider a binary input as proposed in \cite{KHMBK-98}.

The paper is organized as follows.
In Section \ref{sec:encod}, we first describe a coset coding strategy suitable
for lattices, namely using coset lattice codes. The corresponding decoding 
strategy is described in Section \ref{sec:wiretap}. The chore results are
given in Section \ref{sec:cd}: Bob's and Eve's probability of decoding
coset lattice codes are computed, and we show that the behaviour of its 
theta series captures what makes a lattice good for being a
wiretap code, motivating the introduction of the notion of secrecy gain.

%
%
\section{Wiretap Lattice Encoding}
\label{sec:encod}

We consider a Gaussian wiretap channel, namely a broadcast channel where the
source (Alice) sends a signal to a legitimate receiver (Bob), while an
illegitimate eavesdropper (Eve) can listen to the transmission. It is modeled
by
\[
\begin{array}{ccl}
y &= & x + v_b\\
z &= & x + v_e,
\end{array}
\]
where $x$ is the transmitted signal, $v_b$ and $v_e$ denote the Gaussian
noise at Bob, respectively Eve's side, both with zero mean, and respective
variance $\sigma_b^2$ and $\sigma_e^2$.
We assume that Bob has a good $\mathsf{SNR}$, but that 
$\sigma_{b}^{2}=N_{0}<<N_{1}=\sigma_{e}^{2}$, so that Eve has a poor 
$\mathsf{SNR}$, in particular with respect to Bob.

Alice's encoder maps $k$ information symbols $s_1,\ldots,s_k$
from $\Sc=\{0,1\}$ to a codeword $\xv=(x_1,\ldots,x_n)\in\RR^n$, and over a
transmission of $n$ symbols, we get
\begin{equation}\label{eq:wiretap}
\begin{array}{ccl}
\yv &= & \xv + \vv_b\\
\zv &= & \xv + \vv_e.
\end{array}
\end{equation}

Alice uses lattice coding, that is the codeword $\xv=(x_1,\ldots,x_n)$
is actually a lattice point. A lattice $\Lambda$ is a discrete set of points
in $\RR^n$, which can be described in terms of its generator matrix $M$ by 
(\cite{OV-04,CS-98}) 
\[
\Lambda = \{ \xv = \uv M ~|~ \uv \in \ZZ^m \},
\]
where
\[
M = \left( \begin{array}{cccc}
          v_{11}& v_{12} &\ldots &v_{1n} \\
          v_{21}& v_{22} &\ldots &v_{2n} \\
             \ldots &        & \ldots & \\
          v_{m1} & v_{m2}& \ldots & v_{mn} \\
           \end{array}  \right)
\]
and 
\begin{eqnarray*}
\vv_1 & = & (v_{11},v_{12},\ldots,v_{1n}), \\
\vv_2 & = & (v_{21},v_{22},\ldots,v_{2n}), \\
      & \ldots & \\
\vv_m &= & (v_{m1},v_{m2},\ldots,v_{mn}),
\end{eqnarray*}
are a linearly independent set of vectors in $\RR^n$ (so that $m\leq n$) 
which form a basis of the lattice.

Alice chooses a lattice $\Lambda_b$ (we use the subscript $b$ to refer to the 
intended legimitate receiver Bob) and then encodes her $k$ bits of information 
into a point $\xv\in\Lambda_b$:
\[
\sv=(s_1,\ldots,s_k)\in \{0,1\}^k \mapsto \xv=(x_1,\ldots,x_n)\in \Lambda_b.
\]
In a practical scenario, a finite subset of $\Lambda_b$ must be chosen as a
function of the available power at the receiver, though for the analysis,
we will often consider the infinite lattice, which is simpler to understand
since we do not need to take into account the boundary effect.

In order to get confusion at the eavesdropper, we use coset coding, as proposed
in \cite{W-75,OW-84}. The idea (which has been used ever since, whenever there
is wiretap coding) is that instead of having a one-to-one correspondence
between $\sv\in\{0,1\}^k \leftrightarrow \xv\in\Lambda_b$, the vector of
information symbols is mapped to a set of codewords, namely a coset (that is, 
a set of points obtained by translation of a lattice), after
which a random point to be actually transmitted is chosen randomly inside the
coset. More precisely, we partition the lattice $\Lambda_b$ into a union of
disjoint cosets of the form
\[
\Lambda_e+\cv,
\]
with $\Lambda_e$ a sublattice of $\Lambda_b$ and $\cv$ an $n$-dimensional
vector not in $\Lambda_e$. We need $2^k$ cosets to be labelled by 
$\sv\in\{0,1\}^k$:
\[
\Lambda_b = \cup_{j=1}^{2^k} (\Lambda_e+\cv_j).
\]
Since every coset contains the same number of elements, we have that
\begin{equation}\label{eq:nbcosets}
\left| \Lambda_b/\Lambda_e \right|=2^k.
\end{equation}
Once the mapping
\[
\sv \mapsto \Lambda_e+\cv_{j(\sv)}
\]
is done, Alice randomly chooses a point $\xv\in \Lambda_e+\cv_{j(\sv)}$ and
sends it over the wiretap channel. This is equivalent to choose a random 
vector $\rv\in\Lambda_e$. The transmitted lattice point $\xv\in\Lambda_b$ is 
finally of the form
\begin{equation}\label{eq:xsent}
\xv = \rv + \cv \in \Lambda_e + \cv.
\end{equation}
We have denoted the sublattice $\Lambda_e$,
since it encodes the random bits that are there to increase Eve's confusion,
and is then the lattice intended for Eve.

\begin{example}\rm
Take $\Lambda_b=\ZZ^2$ in $\RR^2$ and $\Lambda_e=2\ZZ^2$, for
which we have
\begin{eqnarray*}
\ZZ^2\!\!\!\! &=& \!\!\!\{(x,y),~x,y\in\ZZ\} \\
 \!\!\!\!&=& \!\!\!2\ZZ^2 \cup (2\ZZ^2+(0,1)) \cup (2\ZZ^2+(1,0))
\cup (2\ZZ^2+(1,1)).
\end{eqnarray*}
The lattice $\ZZ^2$ is thus partionned into $2^k=4$ cosets, allowing to
transmit $k=2$ bits of information.
Alice can then label any of the above 4 cosets, say
\[
\begin{array}{ll}
00 \mapsto 2\ZZ_2,&01 \mapsto (2\ZZ_2+(0,1)),\\
10 \mapsto (2\ZZ_2+(1,0)),&11 \mapsto (2\ZZ_2+(1,1)).
\end{array}
\]
To transmit the two bits $01$, she then randomly picks a point in the coset
$2\ZZ_2+(0,1)$, say $\xv=(2,3)$, that is
\[
\xv = \rv + \cv = 2(1,1) + (0,1),
\]
and sends this point over the wiretap channel.
\end{example}

By using lattice coset encoding, we notice that two lattices play a role:
\begin{itemize}
\item
the lattice $\Lambda_b$, that Alice uses to communicate reliably with Bob,
\item
the lattice $\Lambda_e$, which is a sublattice of $\Lambda_b$, that appears
in the process of coset coding for encoding random bits.
\end{itemize}

Our goal is to study how the properties of these two lattices are related
to the design of good wiretap codes.

%
%
\section{Wiretap Lattice Decoding}
\label{sec:wiretap}

After transmission over the Gaussian wiretap channel, Bob and Eve receive 
respectively (see (\ref{eq:wiretap}) and (\ref{eq:xsent}))
\[
\begin{array}{ccll}
\yv &= & \xv + \vv_b &= \rv + \cv + \vv_b\\
\zv &= & \xv + \vv_e &= \rv + \cv + \vv_e,
\end{array}
\]
where we recall that $\rv\in\Lambda_e$ encodes the random bits, and
$\cv$ is the coset representative of minimum energy labelled by the 
information bits.
Both Bob and Eve are interested in decoding the information bits, namely
in finding the correct coset that was sent. To do so, they need to find the
closest lattice point in $\Lambda_b$ to their respective received signal 
$\yv$ or $\zv$, from which they deduce the coset to which it corresponds.

Recall that for any lattice point $P_i$ of a lattice $\Lambda\subset \RR^n$,
its Voronoi cell is defined by
\[
\Vc(P_i)=\{\xv\in \RR^n,~d(\xv,P_i)\leq d(\xv,P_j)\mbox{ for all }j\}.
\]
Since all lattice points have the same Voronoi cell, we will speak of the
Voronoi cell of the lattice $\Lambda$ and denote it by $\Vc(\Lambda)$.

Now when transmitting a codeword $\xv_k$ in $\RR^n$ with Voronoi cell
$\Vc(\xv_k)$ over an additive white Gaussian noise channel with noise variance 
$\sigma^2$, the decoder makes the correct decision if and only if the noise 
vector is in $\Vc(\xv_k)$, an event of probability
\[
\frac{1}{(\sigma\sqrt{2\pi})^n}\int_{\Vc(\xv_k)}e^{-||\uv||^2/2\sigma^2}d\uv.
\]
In our scenario, the probability $P_c$ of correct decision concerns not just 
one point but a coset, and thus it is the probability that the received signal
lies in the union of the Voronoi regions of $\Lambda_b$, translated by points
of $\Lambda_e$.
Suppose that the lattice point $\xv_k=\rv_k+\cv_k\in\Lambda_b$ has been 
transmitted. The probability $P_c$ of finding the correct coset is thus,
assuming no boundary effect
\[
P_c=\frac{1}{(\sigma\sqrt{2\pi})^n}
\sum_{\rv\in\Lambda_e}\int_{\Vc(\xv_k)+\rv}e^{-||\uv||^2/2\sigma^2}d\uv.
\]

If we take $M$ codewords $\xv_1,\ldots,\xv_M$ from $\Lambda_b$, then as 
already noticed, all Voronoi cells are the same, namely 
$\Vc(\xv_k)=\Vc(\Lambda)$, $k=1,\ldots,M$, and thus we get
\begin{equation}\label{eq:Pe}
P_c=\frac{1}{(\sigma\sqrt{2\pi})^n}
\sum_{\rv\in\Lambda_e}\int_{\Vc(\Lambda_b)+\rv}e^{-||\uv||^2/2\sigma^2}d\uv.
\end{equation}

%
%

\section{Wiretap Lattice Code Design}
\label{sec:cd}

We now study the probability of Bob and Eve to make a correct decoding 
decision, and try to maximize Bob's probability while minimizing the one of 
Eve. This leads us to study the theta series of the lattices involved.

\subsection{A first analysis}

Considering the wiretap channel (\ref{eq:wiretap}) where Alice transmits
lattice codewords from an $n$-dimensional lattice $\Lambda_b$, we thus get 
from (\ref{eq:Pe}) that the probability $P_{c,b}$ of Bob's
(resp. $P_{c,e}$ of Eve's) correct decision is:
\begin{eqnarray*}
P_{c,b}&=&\frac{1}{(\sqrt{2\pi}\sigma_b)^n}\sum_{\rv\in\Lambda_e}
\int_{\Vc(\Lambda_b)+\rv}e^{-\Vert \uv\Vert ^2/2\sigma_b^2}d\uv\\
P_{c,e}&=&\frac{1}{(\sqrt{2\pi}\sigma_e)^n}\sum_{\rv\in\Lambda_e}
\int_{\Vc(\Lambda_b)+\rv}e^{-\Vert \uv\Vert ^2/2\sigma_e^2}d\uv.
\end{eqnarray*}

Since by assumption Bob has a good $\mathsf{SNR}$, its received vector
$\yv$ is most likely to lie in the Voronoi region around the origin,
and thus the terms corresponding to $\rv\neq {\bf 0}$ in (\ref{eq:Pe})
are negligeable, which yields:
\begin{equation}\label{eq:Pcb}
P_{c,b}\simeq\frac{1}{(\sqrt{2\pi}\sigma_b)^n}
\int_{\Vc(\Lambda_b)}e^{-\Vert \uv\Vert ^2/2\sigma_b^2}d\uv.
\end{equation}
This is now the familiar case of transmitting lattice points over the
Gaussian channel, for which it is known that $\Lambda_b$ should have a good
Hermite parameter, to get a good coding gain.

We are on the contrary under a low $\mathsf{SNR}$ assumption for Eve, namely
$\sigma_{e}$ is large, and thus a Taylor expansion at order 0 gives
\[
e^{-||\wv+\rv||^2/2\sigma^2}=e^{-||\rv||^2/2\sigma^2}
+O\left(\frac{1}{\sigma_e^2} \right)
\]
so that
\begin{align*}
\int_{\Vc(\Lambda_b)+\rv}e^{-||\uv||^2/2\sigma^2}d\uv
&=  &\int_{\Vc(\Lambda_b)}e^{-||\wv+\rv||^2/2\sigma^2}d\wv \\
&\simeq  & \vol(\Vc(\Lambda_b))e^{-||\rv||^2/2\sigma^2},
\end{align*}
where the volume $\Vc(\Lambda_b)$ of the lattice is
\[
\vol(\Vc(\Lambda))=\int_{\Vc(\Lambda)} d\xv =\det(MM^T)^{1/2}.
\]

The probability of making a correct decision for Eve is then
\begin{equation}\label{eq:Pce}
P_{c,e}\simeq\frac{1}{(\sqrt{2\pi}\sigma_e)^n}\vol(\Vc(\Lambda_b))
\sum_{\rv\in\Lambda_e}e^{-\Vert \rv\Vert ^2/2\sigma_e^2},
\end{equation}
from which we get that
\begin{equation}\label{eq:ratio}
\frac{P_{c,e}}{P_{c,b}}\simeq\left(\frac{\sigma_{b}}{\sigma_{e}}\right)^n
\vol(\Vc(\Lambda_b))\frac{
\sum_{\rv\in\Lambda_e}e^{-\Vert \rv\Vert ^2/2\sigma_e^2}}
{
\int_{\Vc(\Lambda_b)}e^{-\Vert \uv\Vert ^2/2\sigma_b^2}d\uv.
}
\end{equation}

We know how to design good codes for Bob's channel, and have
his probability of making a correct decision arbitrarily close to 1.
Our aim is thus to minimize the probability $P_{c,e}$ of Eve making a correct
decision, while keeping $P_{c,b}$ unchanged.
This is equivalent to minimize (\ref{eq:ratio}), that is to find a lattice 
$\Lambda_{b}$ which is as good as possible for the Gaussian channel 
\cite{CS-98}, and
\begin{equation}
\label{eq:cd}
\boxed{
\begin{array}{c}
\mbox{minimize }\sum_{\rv\in\Lambda_e}e^{-\Vert \rv\Vert ^2/2\sigma_e^2}\\
\mbox{under the constraint } \log_2\left| \Lambda_b/\Lambda_e \right|=k.
\end{array}
}
\end{equation}
The constraint on the cardinality of cosets (or rate) is equivalent
to set the fundamental volume of $\Lambda_e$ equal to a constant. 

It is natural to start by approximating the sum of exponentials by its terms 
of higher order, namely
\begin{align*}
\sum_{\rv\in\Lambda_e}e^{-\Vert \rv\Vert ^2/2\sigma_e^2}
&\simeq& 1+\sum_{\rv\in\Lambda_e,||\rv||=d_{\min}(\Lambda_e)}
e^{-\Vert \rv\Vert ^2/2\sigma_e^2}\\
& = & 1+ \tau(\Lambda_e)e^{-d_{\min}(\Lambda_e)^2/2\sigma_e^2},
\end{align*}
where $\tau(\Lambda_e)$ is the kissing number of $\Lambda_e$ which counts
the number of vectors of length $d_{\min}(\Lambda_e)$.
Thus as a first criterion, we should maximize $d_{\min}(\Lambda_e)$ while 
preserving the fundamental volume of $\Lambda_e$, which is equivalent to 
require for $\Lambda_e$ to have a good Hermite parameter
\[
\gamma_H(\Lambda)=\frac{d_{\min}^2(\Lambda)}{\det(MM^T)^{1/n}}.
\]
after which we should minimize its kissing number.
This approximation however assumes high $\mathsf{SNR}$, which typically 
Eve does not have. We thus cannot be content with this approximation, and 
have to obtain a more precise analysis.

\subsection{The secrecy gain}

Let us get back to the code design criterion (\ref{eq:cd}) and rewrite it in 
terms of the theta serie of the lattice considered. 

Recall that given a lattice $\Lambda \subset\RR^n$, its {\em theta serie} 
$\Theta_\Lambda$ is defined by (\cite{CS-98})
\begin{equation}
\label{def:theta}
\Theta_{\Lambda}(z)=
\sum_{\boldsymbol{x}\in\Lambda}q^{\left\Vert \boldsymbol{x}\right\Vert ^{2}},
~q=e^{i\pi z},\mathrm{Im}(z)>0.
\end{equation}

Exceptional lattices have theta series that can be expressed as functions
of the Jacobi theta functions $\vartheta_{i}(q)$, $i=2,3,4$, themselves 
defined by
\begin{align*}
\vartheta_{2}(q) & =\sum_{n=-\infty}^{+\infty}q^{\left(n+\frac{1}{2}\right)^{2}},\\
\vartheta_{3}(q) & =\sum_{n=-\infty}^{+\infty}q^{n^{2}},\\
\vartheta_{4}(q) & =\sum_{n=-\infty}^{+\infty}\left(-1\right)^{n}q^{n^{2}}.
\end{align*}

\begin{example}
Here are a few examples of theta series for some exceptional lattices. 
\begin{enumerate}
\item
The cubic lattice $\mathbb{Z}^{n}$: 
\[
\Theta_{\ZZ^n}(q)=\vartheta_{3}(q)^{n}.
\]
\item
$D_{n}$: 
\[
\Theta_{\Lambda}(D_n)
=\frac{1}{2}\left(\vartheta_{3}(q)^{n}+\vartheta_{4}(q)^{n}\right).
\]
\item
The Gosset lattice $E_{8}$: 
\[
\Theta_{E_8}(q)=\frac{1}{2}\left(\vartheta_{2}(q)^{8}+\vartheta_{3}(q)^{8}
+\vartheta_{4}(q)^{8}\right).
\]
\item
The Leech lattice $\Lambda_{24}$:
\begin{eqnarray*}
\Theta_{\Lambda_{24}}(q)
&=&\frac{1}{8}\left(\vartheta_{2}(q)^{8}+\vartheta_{3}(q)^{8}
   +\vartheta_{4}(q)^{8}\right)^{3}\\
&& -\frac{45}{16}\vartheta_{2}(q)^{8}\vartheta_{3}(q)^{8}\vartheta_{4}(q)^{8}.
\end{eqnarray*}
\end{enumerate}
\end{example}

From (\ref{eq:cd}), we need to minimize 
\begin{eqnarray*}
\sum_{\rv\in\Lambda_e}e^{-\Vert \rv\Vert ^2/2\sigma_e^2} 
&=&\sum_{\rv\in\Lambda_e} \left( e^{-1/2\sigma_e^2}\right)^{||\rv||^2} \\
&=&\sum_{\rv\in\Lambda_e} \left( (e^{i\pi})^{-1/2i\pi\sigma_e^2}\right)^{||\rv||^2} \\
&=& \Theta_{\Lambda_{e}}\left(z= \frac{-1}{2i\pi\sigma_{e}^{2}} \right)
\end{eqnarray*}
with $q=e^{i\pi z}$ and
\[
{\rm Im}\left(\frac{-1}{2i\pi\sigma_{e}^{2}}\right)=
{\rm Im}\left(\frac{i}{2\pi\sigma_e^2}\right)>0.
\]
Thus to minimize Eve's probability of correct decision is equivalent to
minimize $\Theta_{\Lambda_{e}}(z)$ in $z=i/2\pi\sigma_e^2$. To approach this 
problem, let us set $y=-iz$ and restrict to real positive values of $y$. 
We are now interested in minimizing 
\[
\Theta_{\Lambda_e}(y)=\sum_{\rv\in\Lambda_e}q^{\left\Vert \rv \right\Vert ^{2}},
~q=e^{-\pi y},y>0, 
\]
in the particular value of $y$ corresponding to $z=i/2\pi\sigma_e^2$, namely 
\[
y=\frac{1}{2\pi\sigma_e^2}.
\] 
This is actually a problem that classically arises in the study of theta 
series \cite{C-06}: given the lattice dimension $n$, find the lattice 
$\Lambda^{\star}$ that minimizes $\Theta_{\Lambda}(y)$ for a given value of $y$. 

Note that if $\Lambda_{e}$ is not chosen to be a particular lattice, we can 
assume that $\Lambda_e=\mathbb{Z}^{n}$.
We consequently define the secrecy function of a given lattice $\Lambda$ as 
the ratio of its theta series and the theta series of $\ZZ^n$, in a chosen 
point $y$.
\begin{defn}
Let $\Lambda$ be an $n-$dimensional lattice. The \textit{secrecy function} 
of $\Lambda$ is given by
\[
\Xi_{\Lambda}(y)=\frac{\Theta_{\mathbb{Z}^{n}}(y)}{\Theta_{\Lambda}(y)}
=\frac{\vartheta_{3}(y)^{n}}{\Theta_{\Lambda}(y)}
\]
defined for $y>0$.
\end{defn}

As we want to minimize the expression of Eve's probability of correct decision 
in (\ref{eq:cd}), we are interested in the maximum value of the secrecy 
function. This yields the notion of secrecy gain.
\begin{defn}
The \textit{secrecy gain} $\chi_{\Lambda}$ of an $n-$dimensional lattice 
$\Lambda$ is defined by
\[
\chi_{\Lambda}= \sup_{y>0}\Xi_{\Lambda}(y).
\]
\end{defn}

\begin{figure}
\includegraphics[scale=0.8]{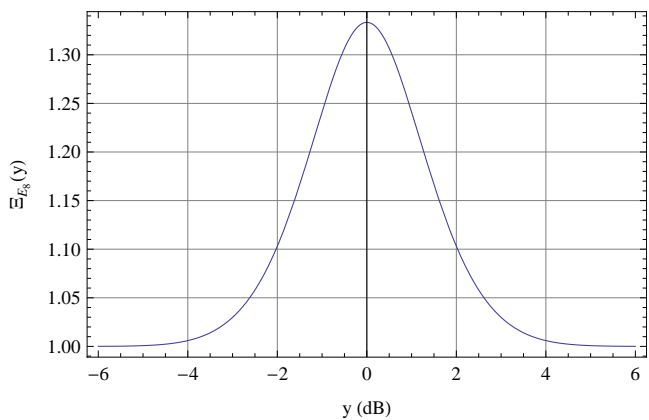}
\caption{
\label{fig:E8}
Secrecy function of $E_8$.}
\end{figure}

\begin{figure}
\includegraphics[scale=0.8]{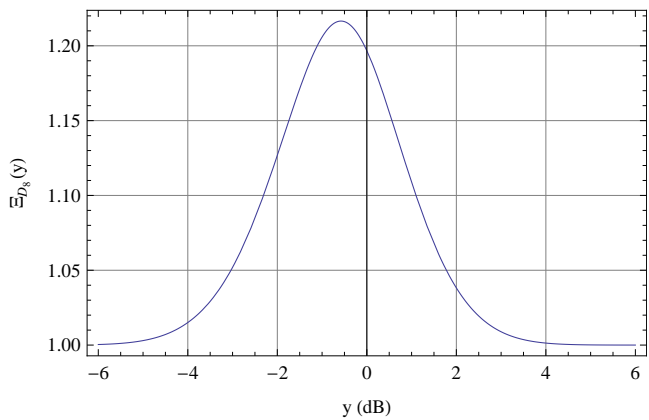}
\caption{
\label{fig:D8}
Secrecy function of $D_8$.}
\end{figure}

Examples of the secrecy function for the lattices $E_8$ and $D_8$ are shown 
in Figures \ref{fig:E8} and \ref{fig:D8}, respectively. Both lattices clearly 
have a maximum, happening in $y=1$ for $E_8$ but not for $D_8$ (which is conjectured to have this maximum in 
$\tfrac{1}{\sqrt[4]{2}}$). 


It is worth emphasizing that the value at which the secrecy function gets 
its maximum is important for the code design, since it tells us what is the 
$\mathsf{SNR}$ at which the wiretap lattice code is providing most confusion 
to Eve. The two examples suggest that this value depends on the chosen lattice.

Conjectures on the behaviour of the secrecy gain are currently being 
investigated. It is expected that an asymptotic analysis will give a first 
insight, and that a finer study should reveal how the secrecy gain is connected 
to the equivocation rate.

Let us conclude by giving a small example of code construction.

\begin{example}\rm
Consider the case of an $8-$dimensional (real) construction. Suppose
we want to transmit at a secrecy rate of $2$ bits per complex symbol.
We choose $\Lambda_{b}=E_{8}$, since this lattice has the best coding gain
and the best shaping gain in dimension $8$. For $\Lambda_{e}$, we
choose as sublattice of $E_8$ the lattice $2E_{8}$, a scaled version of
$\Lambda_{b}$. We then have
\[
\left| E_{8}/2E_{8}\right|=256
\]
which gives as rate per complex symbol
\[
R=\frac{1}{4}\log_{2}\left| E_{8}/2E_{8}\right|=2
\]
which is the requested rate.

To construct $E_{8}$ while preserving the overall shaping, we choose a 
construction $A$ \cite{CS-98}:
\[
E_{8}=2\mathbb{Z}^{8}+(8,4,4)\]
where $(8,4,4)$ stands for the Reed-Müller code of length $8$ and
dimension $4$. We repeat the same construction for
$2E_{8}$, namely
\[
2E_{8}=4\mathbb{Z}^{8}+2\cdot(8,4,4).
\]
We can now give a construction of $E_{8}$ using $2E_{8}$:
\[
E_{8}=2E_{8}+(8,4,4)+2\cdot\mathcal{C}_{\nmid}\]
where $\mathcal{C}_{\nmid}$ is the set of all representatives of
the cosets of $(8,4,4)$ with minimum Hamming weights. This yields\[
E_{8}/2E_{8}=(8,4,4)+2\cdot\mathcal{C}_{\nmid}\]
which gives an alphabet with $256$ codewords.

Then, the random bits label $2E_8$, which means that $4$ of these bits serve 
as information bits for $2\cdot(8,4,4)$ and the other ones label points of 
$4\mathbb{Z}^{8}$.

It has been proved \cite{C-06}, and this is the best result known up to date, 
that some lattices, including $E_8$, reach a local minimum of their theta 
series for some constant $y>0$ close to 1. Thus using $\Lambda_e=2E_8$ indeed 
helps in optimizing the secrecy gain.
 
\end{example}

%
%

\section{Current and future work}

In this paper, we provided a practical wiretap coding scheme using coset
lattice codes. We exhibited geometric properties that a lattice
and its sublattice should satisfy to provide good wiretap codes for 
transmission over additive white Gaussian noise channels, in terms of the 
theta series of the involved lattices. This yielded the notion of secrecy gain. 
Our analysis focuses on error probabilities of both users rather than 
on equivocation rate, though we expect that further work will enlighten the 
connection between the two concepts.

We are currently studying different conjectures on the behaviour of the 
secrecy gain, as well as the design of lattice codes that fullfil
the code design criteria. Having explicit constructions of families of wiretap 
codes to compare will give us a further understanding of what is a good
wiretap lattice code. It is also a natural work to address the achievability 
of such codes with respect to the secrecy capacity of Gaussian channels.

Finally, lattice codes have also been useful to design modulation schemes for 
fading channels. It is a natural generalization to consider a similar analysis 
of what makes a good wiretap code in the context of fading channels.

%
%

\section*{Acknowledgment}

The research of F. Oggier is supported in part by the Singapore National
Research Foundation under Research Grant NRF-RF2009-07 and NRF-CRP2-2007-03,
and in part by the Nanyang Technological University under Research
Grant M58110049 and M58110070.

%
%


\begin{thebibliography}{99}
%
\bibitem{CS-98}
J.H. Conway, N.J.A. Sloane, ``Sphere packings, Lattices and Groups,"
Third edition, {\em Springer-Verlag}, New York, 1998.
%
\bibitem{C-06}
R. Coulangeon, ``Spherical designs and zeta functions of lattices,'' 
{\em http://arxiv.org/abs/math/0611735}.
%
\bibitem{ES-07}
S. Y. El Rouayheb, E. Soljanin, ``On Wiretap Networks II,"
{\em Proc. ISIT}, 2007.
%
\bibitem{HY-09}
Xiang He and Aylin Yener, ``Providing Secrecy With Structured Codes: Tools and Applications to Two-User Gaussian Channels,'' preprint, arxiv.org/pdf/0907.5388.
%
\bibitem{KHMBK-98}
D.Klinc, J.Ha, S.W.McLaughlin, J.Barros, and B.J.Kwak, ``LDPC Codes for the Gaussian Wiretap Channel,''  {\em Proc. Information Theory Workshop (ITW)}, Oct. 2009.
%
\bibitem{OV-04}
F. Oggier, E. Viterbo, ``Algebraic number theory and code design for Rayleigh
fading channels,'' {\em Foundations and Trends
in Communications and Information Theory}, December 2004.
%
\bibitem{OW-84}
L. H. Ozarow and A. D. Wyner,``Wire-tap channel II,'' {\em Bell Syst. Tech.
Journal}, vol. 63, no. 10, pp. 2135-2157, Dec. 1984.
%
\bibitem{PR-05}
S.S. Pradhan, K. Ramchandran, ``Generalized Coset Codes for Distributed 
Binning'', {\em IEEE Transactions on Information Theory}, OCtober 2005.
%
\bibitem{TDCMM-07}
A. Thangaraj, S. Dihidar, A. R. Calderbank, S.W. McLaughlin, and J.-M. Merolla,``Applications of LDPC Codes to the Wiretap Channel,'' {\em IEEE Transactions on Information Theory}, vol. 53, No. 8, Aug. 2007
%
\bibitem{W-75}
A.D. Wyner,``The wire-tap channel,'' {\em Bell. Syst. Tech. Journal},
vol. 54, October 1975.
%
\bibitem{ZSE-02}
R. Zamir, S. Shamai, U. Erez, ``Linear/Lattice Codes for Structured 
Multi-terminal Binning'', {\em IEEE Transactions on Information Theory}, 
June 2002.
%
\end{thebibliography}
\end{document}